\RequirePackage{fix-cm}
\documentclass[twocolumn,epjc3]{svjour3}  
\smartqed  
\pdfoutput=1
\usepackage{graphicx}
\graphicspath{{figures/}}
\usepackage{amsmath}
\usepackage{amsfonts}
\usepackage{amssymb}
\usepackage{multirow}

\usepackage[colorlinks=true, linkcolor=red, citecolor=blue]{hyperref}

\def\({\left(}
\def\){\right)}
\def\[{\left[}
\def\]{\right]}
\def\be{\begin{eqnarray}}
\def\ee{\end{eqnarray}}

\usepackage{acro}
\usepackage{adjustbox}
\usepackage{array}

\newcommand{\GeV}{{\rm GeV}}
\newcommand{\m}{{\rm m}}

\newcommand{\bea}{{\begin{eqnarray}}}
\newcommand{\eea}{{\end{eqnarray}}}

\DeclareAcronym{gw}{
	short = GW ,
	long = gravitational wave ,
	short-plural = s ,
}
\DeclareAcronym{spa}{
	short = SPA ,
	long = stationary phase approximation ,
	short-plural = ,
}
\DeclareAcronym{pn}{
	short = PN ,
	long = post-Newtonian ,
	short-plural = ,
}
\DeclareAcronym{bbh}{
	short = BBH ,
	long = binary black hole ,
	short-plural = s ,
}
\DeclareAcronym{psd}{
	short = PSD ,
	long = power spectral density ,
	short-plural = s ,
}
\DeclareAcronym{snr}{
	short = SNR ,
	long = signal-to-noise ,
	short-plural = ,
}
\DeclareAcronym{ligo}{
	short = LIGO ,
	long = Laser Interferometer Gravitational-Wave Observatory ,
	short-plural = ,
}
\DeclareAcronym{eft}{
	short = EFT ,
	long = effective field theory ,
	short-plural = ,
}
\DeclareAcronym{liv}{
	short = LIV ,
	long = Lorentz invariance violation ,
	short-plural = ,
}
\DeclareAcronym{qg}{
	short = QG ,
	long = quantum gravity ,
	short-plural = ,
}
\DeclareAcronym{sme}{
	short = SME ,
	long = Standard Model Extension ,
	short-plural = ,
}
\DeclareAcronym{cbc}{
	short = CBC ,
	long = compact binary coalescence ,
	short-plural = s ,
}
\DeclareAcronym{grb}{
	short = GRB ,
	long = gamma-ray burst ,
	short-plural = s ,
}
\DeclareAcronym{gr}{
	short = GR ,
	long = general relativity ,
	short-plural =  ,
}
\DeclareAcronym{pdf}{
	short = PDF ,
	long = probability distribution function ,
	short-plural = s ,
}

\journalname{Eur. Phys. J. C}
\begin{document}

\title{Tests of CPT invariance in gravitational waves with LIGO-Virgo catalog GWTC-1
}


\author{Sai Wang\thanksref{addr1,addr2}
        \and
        Zhi-Chao Zhao\thanksref{e1,addr3} 
}

\thankstext{e1}{Correspondence author: zhaozc@bnu.edu.cn}


\institute{Theoretical Physics Division, Institute of High Energy Physics, Chinese Academy of Sciences, Beijing 100049, People's Republic of China \label{addr1}
           \and
           School of Physical Sciences, University of Chinese Academy of Sciences, Beijing 100049, People's Republic of China \label{addr2}
           \and
           Department of Astronomy, Beijing Normal University, Beijing 100875,  People's Republic of China \label{addr3}
}

\date{Received: date / Accepted: date}

\maketitle

\begin{abstract}
A discovery of gravitational waves from binary black holes raises a possibility that measurements of them can provide strict tests of CPT invariance in gravitational waves. When CPT violation exists, if any, gravitational waves with different circular polarizations could gain a slight difference in propagating speeds. Hence, the birefringence of gravitational waves is induced and there should be a rotation of plus and cross modes. For CPT-violating dispersion relation ${\omega^{2}=k^{2}}$ ${\pm 2\zeta k^{3}}$, where a sign ${\pm}$ denotes different circular polarizations, we find no substantial deviations from CPT invariance in gravitational waves by analyzing a compilation of ten signals of binary black holes in the LIGO-Virgo catalog GWTC-1. 
We obtain a strict constraint on the CPT-violating parameter, i.e., $\zeta=0.14^{+0.22}_{-0.31}\times10^{-15}\text{m}$, which is around two orders of magnitude better than the existing one. Therefore, this study stands for the up-to-date strictest tests of CPT invariance in gravitational waves.
\end{abstract}

\section{Introduction}
CPT invariance \cite{Peskin:1995ev}, which is a simultaneous reversal of charge, parity and time, is well known as one of the fundamental laws of physics. 
Since it was proposed, it has been tested with high precision by a variety of observations in laboratories and astronomy \cite{Kostelecky:2008ts,Will:2014kxa}. 
The measurements of the net polarization of \aclp{grb} have displayed the strictest testes of CPT invariance in the pure photon sector \cite{Toma:2012xa,Myers:2003fd,Amelino-Camelia:2016ohi}. 
There were also strict tests in the pure neutrino sector \cite{Amelino-Camelia:2016ohi,Jacob:2006gn}. 
However, few constraints have been placed on possible deviations from CPT invariance in the pure gravitational sector. 
Theoretically, the \acl{qg} at Planck scale $\sim10^{19}\GeV$ is expected to leave low-energy relic effects \cite{Kostelecky:1988zi,AmelinoCamelia:1997gz,AmelinoCamelia:2008qg,Mielczarek:2017cdp}, wherein CPT violation is one famous example. 
Since CPT invariance is a fundamental law in nature, it is well justified to be unstinting in one's efforts to explore CPT violation under various circumstances.

The discovery of \ac{bbh} coalescences by the \ac{ligo} \cite{Abbott:2016blz} opens a clean observational window to provide strict tests of fundamental physics \cite{Isi:2019asy,LIGOScientific:2019fpa,Isi:2019aib,Yunes:2016jcc,Addazi:2018uhd,Wu:2016igi,Cardoso:2017cqb,Ezquiaga:2018btd,Wang:2016ana,Chang:2019xcb}. Here, we perform such a test of CPT invariance in \acp{gw}. 
When CPT invariance was deviated, if any, \acp{gw} with left-handed and right-handed circular polarizations could gain a slight difference in propagating speeds \cite{Zhao:2019xmm,Kostelecky:2016kfm}. 
As a consequence, the birefringence is induced and there is a rotation of plus ($+$) and cross ($\times$) modes \cite{Wang:2017igw,Mewes:2019dhj}. 
The birefringent effect depends on the \ac{gw} frequency and can be accumulated along the trajectory of \acp{gw}, which are emitted from \aclp{cbc} at cosmological distances.
Therefore, we can perform measurements of the polarizations to test CPT invariance in \acp{gw} or detect possible deviations from it. 

This study aims at testing the CPT invariance in \acp{gw} and placing strict limits on the leading-order CPT violation in \acp{gw}. 
Higher-order CPT violations are ignored since they are expected to lead smaller effects in the spirits of \ac{eft} \cite{Kostelecky:2003fs}. 
In this work the CPT-violating dispersion relation is manifested as
\bea
\label{eq:cptdispersion}
\omega^{2}=k^{2}\pm 2\zeta k^{3}\ ,
\eea
where a sign $\pm$ takes $+/-$ for the left-/right-handed circular polarization and $\zeta$ is a length-dimensional parameter characterizing the size of CPT violating effect. 
We assume a convention {\small $\hbar=1$}.  
Hence, $\omega$ and $k$ denote the energy and momentum of gravitons, respectively.
Eq.~(\ref{eq:cptdispersion}) could be related to a dimension-5 CPT-violating operator {\small $\textrm{\r k}_{(V)}^{(5)}$} \cite{Kostelecky:2016kfm}, which is of leading order that causes the birefringence of \acp{gw}. We note {\small $\zeta\simeq\textrm{\r k}_{(V)}^{(5)}$} here.

Laboratory experiments in the non-relativistic limit are insensitive to such a dimension-5 CPT-violating operator, since Newton's law remains unchanged by it \cite{Bailey:2014bta}.
However, \acp{gw} from cosmologically distant \acp{bbh} provide a potential approach to measure it \cite{Kostelecky:2016kfm,Wang:2017igw,Mewes:2019dhj}.
In the pure gravity sector, the only existing upper limit on the dimension-5 CPT-violating operator (absolute value) was reported to be less than $2\times10^{-14}\m$ \cite{Kostelecky:2016kfm}, which was obtained by measuring the width of the peak at the maximal amplitude of GW150914 \cite{Abbott:2016blz}.
It stands for a $\mathcal{O}(0.01)\text{GeV}$ test of CPT invariance in \acp{gw}.

We would adopt Bayesian parameter inferences \cite{Thrane:2018qnx} and obtain stricter limits on CPT violation in this paper. 
Due to CPT violation, the plus and cross modes of \acp{gw} rotate along the propagating direction.
Instead of the net polarization, the knowledge of gravitational waveform is essential to our data analysis, since a method named as matched filtering \cite{Maggiore:1900zz} is used for extracting the signals. 
Fortunately, in \ac{gr}, one can predict exactly the coalescing process of a compact binary system as well as the waveform generated during the process \cite{Lehner:2001wq}. 
We assume that the CPT-violating effect is minimal and it can be ignored at the source \cite{LIGOScientific:2019fpa}. 
The source effect is believed to be far smaller than the leading/\ac{gr} term and appear at a time scale of radiation reaction that is much shorter than the propagating time of \acp{gw}. 
It is thus probably negligible compared with the propagation effect accumulated along a cosmologically distant trajectory. 
In fact, to proceed a self-consistent study, one need use the technique of numerical relativity, which is tricky for the current work. 
However, our study stands for the first step towards a complete understanding of CPT violation in \acp{gw}. 
It can put valuable physical insights on this field and provide helpful guides to future numerical-relativity investigations. 
Given the above assumption, to detect possible deviations from CPT invariance, we still need knowledges of the CPT-violating propagation effect on gravitational waveform. It has been done in Refs.\cite{Wang:2017igw,Mewes:2019dhj}.

The remainder of this work is arranged as follows.
In section \ref{sec:theory}, we introduce the gravitational waveform which is corrected by CPT violation.
In section \ref{sec:method}, we demonstrate the method of data analysis. 
In section \ref{sec:result}, we obtain new constraints on the CPT-violating parameter.
Our conclusions are given in section \ref{sec:last}.

\section{Theory}\label{sec:theory}
\vspace{-0.06cm}

We could evaluate the CPT-violating contribution to the gravitational waveform \cite{Wang:2017igw,Mewes:2019dhj}. 
The eigenstates are consist of two circular polarizations. 
Based on Eq.~(\ref{eq:cptdispersion}), the phase speed of the left-handed circular polarization is given as $v_{\mathrm{L}}\simeq1-\zeta \omega$.
To first order, we obtain the gravitational strain as $h_{\mathrm{L}}(t)\sim \mathrm{e}^{-i\omega(t-l/v_{\mathrm{L}})}\simeq \mathrm{e}^{i\zeta\omega^{2} l}\mathrm{e}^{-i\omega(t-l)}$, where $t$ and $l$ denote the time and distance to the source, respectively.
Hence, the phase is shifted by $\zeta\omega^{2} l$.
Due to the expansion of the universe, we should take the redshift of the energy into account by considering an infinitesimal change in the phase, namely, $d\Psi_{\mathrm{tot}}=d\Psi^{\mathrm{GR}}+\zeta\omega^{2}dl$.
Upon an integration from the source to the detector, the former part gives the phase predicted by \ac{gr}, while the later one gives the correction term due to modified dispersion, i.e., $\delta\Psi=\int \zeta\omega^{2}dl$.
We can replace $dl$ with $dt=-dz/[(1+z)H(z)]$ at zeroth order and multiply $\omega$ by a factor $(1+z)$.
Here, $H(z)$ denotes the Hubble parameter at the redshift $z$.
Therefore, we obtain a finite change in the phase, i.e.,  
\bea\label{eq:deltapsi}
\delta\Psi = 4\pi^2 \zeta f^{2} \int_{0}^{z_{\text{BBH}}}{(1+z)}/{H(z)}dz\ ,
\eea
where $f=\omega/2\pi$ denotes the \ac{gw} frequency in the observer frame, and $z_{\text{BBH}}$ denotes the redshift of a \ac{bbh} in this paper.
The integration over the redshift reveals that the CPT-violating effect accumulates with an increase of cosmological distance.
For the right-handed circular polarization, the phase speed is given as $v_{\mathrm{R}}\simeq 1+\zeta \omega$ and the finite change in the phase becomes $-\delta\Psi$.

\begin{figure}
	\includegraphics[width=1.1\columnwidth]{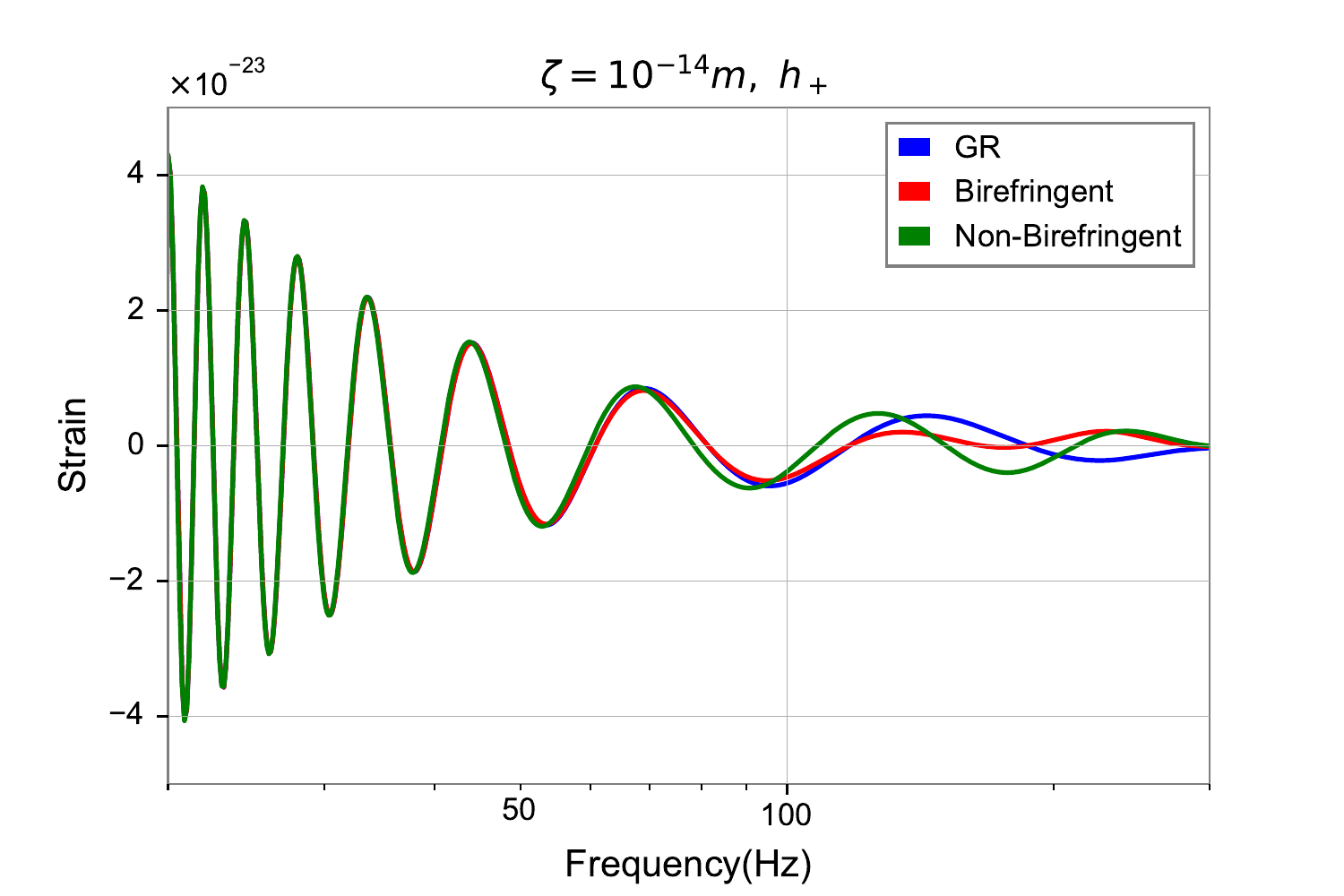}
	\caption{Birefringent gravitational waveform $h_+$ in frequency domain. We let the CPT-violating parameter be $\zeta=10^{-14}\text{m}$. The GR and non-birefringent waveforms are depicted for comparison. [Color online] }
	\label{fig:1}
\end{figure}

Due to CPT violation, the gravitational waveform for circular polarization states is given as
$h_{\mathrm{L,R}}=h_{\mathrm{L,R}}^{\mathrm{GR}}~\mathrm{e}^{\pm i\delta\Psi}$,
where $h^{\mathrm{GR}}$ denotes the waveform predicted by \ac{gr} and $\delta\Psi$ is explicitly given by Eq.~(\ref{eq:deltapsi}).
The circular polarizations are usually decomposed into the plus and cross modes, namely, $h_{\mathrm{L,R}}=h_{+}\pm ih_{\times}$ and $h_{\mathrm{L,R}}^{\mathrm{GR}}=h_{+}^{\mathrm{GR}}\pm ih_{\times}^{\mathrm{GR}}$.
Through a few algebraic operation, therefore, we can represent the CPT-violating waveform $h_{+,\times}$ as a rotation of the CPT-invariant waveform, namely,
\begin{equation}
\label{eq:waveform}
\left(\begin{array}{c} h_{+} \\ h_{\times} \end{array}\right)=\left(\begin{array}{cc} \cos(\delta\Psi) & -\sin(\delta\Psi) \\ \sin(\delta\Psi) & \cos(\delta\Psi) \end{array}\right) \left(\begin{array}{c} h_{+}^{\mathrm{GR}} \\ h_{\times}^{\mathrm{GR}} \end{array}\right)\ .
\end{equation}
For $h^{\mathrm{GR}}$, we adopt the \texttt{IMRPhenomPv2} waveform \cite{Hannam:2013oca,Schmidt:2014iyl}. 
Based on Eq.~(\ref{eq:waveform}), we should note that $\delta\Psi$ is twice the rotation angle due to CPT violation. 
The \ac{gr} waveform would be recovered when we take $\zeta=0$.
Furthermore, the gravitational strain on a given detector is \cite{Maggiore:1900zz}
\be
h=F_{+}h_{+}+F_{\times}h_{\times} \ , 
\ee
where $F_{+,\times}$ denote a set of response pattern functions for the detector such as \ac{ligo}-Hanford, \ac{ligo}-Livingston and Virgo \cite{LSC-Virgo-2019}.
In Fig.~\ref{fig:1}, we plot the birefringent gravitational waveform $h_+$ in frequency domain. 
The CPT-violating parameter is set to be $\zeta=10^{-14}\text{m}$. 
For comparison, we also depict the GR and non-birefringent waveforms, the latter of which is got by simply replacing ``$\pm$'' with ``$+$'' in Eq.~(\ref{eq:cptdispersion}). 
In fact, the non-birefringent case has been extensively studied in Ref.~\cite{LIGOScientific:2019fpa}.
Based on Fig.~\ref{fig:1}, we show that CPT violation can be quantitatively distinguished from other effects. 

\section{Method}\label{sec:method}
For the first time, we perform a parallel Bayesian analysis software. i.e. \texttt{pBilby} \cite{Smith:2019ucc}, to estimate the posterior \acp{pdf} of the CPT-violating parameter $\zeta$ and fifteen binary parameters. We consider a compilation of ten signals of \ac{bbh} coalescences which were reported in the LIGO-Virgo catalog GWTC-1 \cite{LIGOScientific:2018mvr}.
Since the CPT violating effect is expected to be small, a uniform prior \ac{pdf} of $\zeta$ is set as $[-4,4]\times10^{-14}\mathrm{m}$, which is proved to be wide enough for our purpose in the following. Other independent parameters have prior \acp{pdf} matched with Ref.~\cite{LIGOScientific:2019fpa}.

The log-likelihood for a signal with Gaussian noise is defined as \cite{Thrane:2018qnx}
\be
\log\mathcal{L}=\langle s,h(\theta) \rangle-\frac{1}{2}\langle h(\theta),h(\theta) \rangle\ ,
\ee
where $s$ denotes a \ac{gw} signal and $h(\theta)$ denotes a waveform template with parameter space $\theta$. 
An inner product is defined as 
\be
\langle a,b \rangle=4\Re \int_{0}^{\infty}\frac{a(f)b^\ast(f)}{S_{n}(f)}df \ ,
\ee
where $S_{n}(f)$ denotes a single-sided \ac{psd} of detector noise and $\Re$ means the real part.
The waveform template modified by CPT violation is given by Eq.~(\ref{eq:deltapsi}) and Eq.~(\ref{eq:waveform}). 
For each signal in the GWTC-1, we analyze the data collected by detectors which responded to the signal. 
In addition, we employ the noise \acp{psd} of corresponding detectors \cite{LSC-Virgo-2019}.
For multiple detectors, the uncorrelated noises are assumed and the likelihoods should be multiplied together. 
To check the correctness of our method, we reproduce  the results of Table~III in Ref.\cite{LIGOScientific:2018mvr}, without considering the effects of CPT violation.

\section{Results and Discussion}\label{sec:result}
\begin{figure}
	\includegraphics[width=1.1\columnwidth]{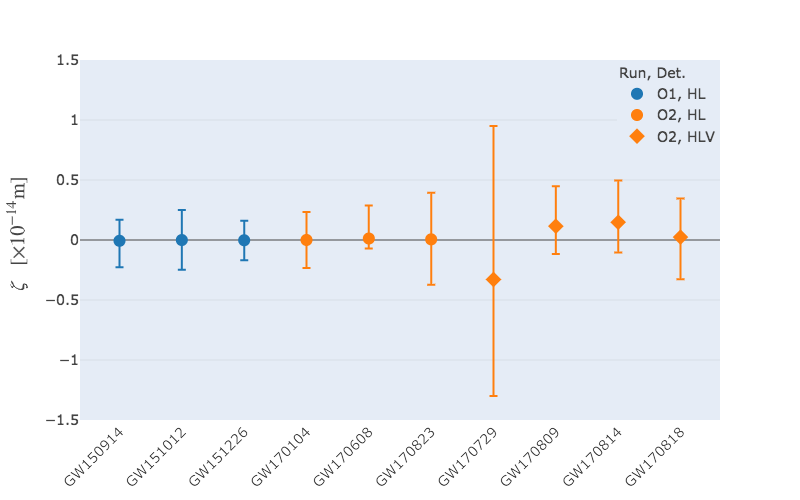}
	\caption{Constraints on the CPT-violating parameter $\zeta$ from the ten signals of \acp{bbh} in the LIGO-Virgo catalog GWTC-1 \cite{LIGOScientific:2018mvr}. A capital H/L/V denotes a \ac{ligo}-Hanford/\ac{ligo}-Livingston/Virgo detector. A dot/square denotes HL/HLV. Blue/orange color denotes O1/O2. The error bars denote 90\% confidence interval. [Color online] }
	\label{fig:2}
\end{figure}

The results of this study are showed as follows.
Fig.~\ref{fig:2} and Tab.~\ref{tab:1} show the strict observational constraints on the CPT-violating parameter $\zeta$ at 90\% confidence level from the ten signals of \acp{bbh} in GWTC-1. 
Since the constraints are well compatible with $\zeta=0$, we find no substantial deviations from CPT invariance, indicating stringent upper limits on $|\zeta|$. 
Typically, we obtain $|\zeta|\lesssim\mathrm{few}\times10^{-15}\mathrm{m}$.
The most stringent constraint is revealed as $\zeta=-0.00^{+0.16}_{-0.17}\times10^{-14}\mathrm{m}$, which is given by GW151226 \cite{Abbott:2016nmj}.
By transforming the posterior \ac{pdf} of $\zeta$ to that of $|\zeta|$, we find that the upper limit on $|\zeta|$ for GW151226 is at least one order of magnitude better than the only existing limit $\lesssim2\times10^{-14}\m$ \cite{Kostelecky:2016kfm}, which was given by GW150914 \cite{Abbott:2016blz}. 
Even for GW150914 itself, our constraint is still one order of magnitude better. 
For other events, we also obtain the stricter constraints than the existing one. 

We can combine the posterior \acp{pdf} of the ten events and obtain a more stringent limit on $\zeta$ than that from an individual event. 
By using \texttt{Monte Python} \cite{Audren:2012wb}, we perform a detailed analysis and obtain  
\be
\zeta=0.14^{+0.22}_{-0.31}\times10^{-15}\text{m}\ ,
\ee
at $1\sigma$ confidence level. 
This limit becomes $0.14^{+0.70}_{-0.56}\times10^{-15}\text{m}$ and $0.14^{+0.94}_{-0.69}\times10^{-15}\text{m}$ at $2\sigma$ and $3\sigma$  confidence levels, respectively. 
Indeed, it is tighter than any individual event. 
Therefore, we obtain the up-to-date strictest constraints on CPT violation in \acp{gw}. 
In addition, our results stand for the first self-consistent test of CPT invariance in \acp{gw}. 
Moreover, it is interesting to compare our method and results with those in a similar work \cite{Shao:2020shv}, which constrained the anisotropic birefringence with the GWTC-1. 
Different from our method, the author utilized the posterior parameter samples released by \ac{ligo}-Virgo Collaborations. He also obtained a combined limit on the CPT-violating parameter, i.e., {$|k_{(V)00}^{(5)}|<3.3\times10^{-16}\mathrm{m}$} at $1\sigma$ confidence level. 
This limit is compatible with ours, since the abnormal propagation of \acp{gw} due to CPT violation, if any, should reside in the residual uncertainty. 
However, it is about three times stronger than ours, since the author followed the different method mentioned above and used all of the GWTC-1 events including GW170817.

\begin{table}[!htp]
	\centering
	\renewcommand{\arraystretch}{1}
	\begin{adjustbox}{angle=0}
		\begin{tabular}{|c|r|}
			\hline
			\textrm{BBH Events} &                $ \zeta~[\times10^{-14}\mathrm{m}]$ \\
			\hline
			GW150914 &   $-0.01^{+0.18}_{-0.22}$ \\
  			GW151012 &   $0.00^{+0.25}_{-0.25}$ \\
			GW151226 &   $0.00^{+0.16}_{-0.17}$ \\
			GW170104 &   $0.00^{+0.23}_{-0.23}$ \\
			GW170608 &   $0.01^{+0.27}_{-0.08}$ \\
			GW170729 &   $-0.33^{+1.28}_{-0.97}$ \\
			GW170809 &   $0.12^{+0.33}_{-0.23}$ \\
			GW170814 &   $0.15^{+0.35}_{-0.25}$ \\
			GW170818 &   $0.02^{+0.32}_{-0.35}$ \\
			GW170823 &   $0.01^{+0.39}_{-0.38}$ \\
			\hline
		\end{tabular}
	\end{adjustbox}
	\caption{Same caption as Fig.~\ref{fig:2}. Typically, we report the stringent constraints as $|\zeta|\lesssim\mathrm{few}\times10^{-15}\mathrm{m}$. The strictest limit is given by GW151226 \cite{Abbott:2016nmj}. The uncertainties denote 90\% confidence interval. }
	\label{tab:1}
\end{table}

It is interesting to qualitatively explore the sources of the observational uncertainties of $\zeta$. 
Naively speaking, we expect a better limit or smaller uncertainty from a more distant \ac{bbh}, since the CPT-violating effect accumulates with the increase of cosmological distance according to Eq.~(\ref{eq:deltapsi}). 
Indeed, it is roughly true for \acp{bbh} with the same chirp mass. 
However, a full story should consider the chirp mass, which determines a cutoff frequency of the signal. 
A \ac{bbh} system with smaller chirp mass could generate on the detectors a temporally longer signal, which is important for an efficient extraction of the CPT-violating effect. 
The side effect is a smaller \ac{snr} \cite{Maggiore:1900zz}, leading larger uncertainties of other parameters.
Therefore, the total uncertainty of $\zeta$ mainly depends on the chirp mass and cosmological distance, as well as the uncertainties of them.
A longer distance or lighter chirp mass is good for the accumulation of the CPT-violating propagation effect, while can reduce \ac{snr}. 

The above discussion is also applicable to GW170817, which is the first detected binary neutron stars.
Though it is lightest, GW170817 is the loudest signal in GWTC-1, since it is nearest. 
In fact, we could obtain a stricter constraint on $\zeta$ from GW170817, following our method. 
However, we would not consider this event in this work. 
Since neutron stars are composed of matter, such as neutrons, it is not pure-gravitational and may involve unknown matter effects. 
To be conservative, therefore, we only focus on tests of CPT invariance with \acp{bbh}. 

\section{Conclusions}\label{sec:last}
In this work, we have shown a systematic test of CPT invariance in \acp{gw}. 
We demonstrated that CPT violation induces the rotation of plus and cross modes of \acp{gw}. 
We showed that the measurements of \acp{gw} from \acp{bbh} provide a clean observational window to test CPT invariance, since only the different circular polarizations are involved.
We performed Bayesian parameter inferences over the ten signals of \acp{bbh} in GWTC-1, but reported no substantial deviations from CPT invariance in \acp{gw}. 
The strictest limit on the CPT-violating parameter was given by GW151226, i.e. $\zeta=0.00^{+0.16}_{-0.17}\times10^{-14}\mathrm{m}$, which is one order of magnitude better than the only existing limit. 
Combining the results of all the ten events, the joint constraint on $\zeta$ was shown to be further improved, i.e., $\zeta=0.14^{+0.22}_{-0.31}\times10^{-15}\text{m}$, which stands for the up-to-date strictest test of CPT invariance in \acp{gw}. 

Our study represents the first self-consistent Bayesian constraints on CPT violation in \acp{gw}, though similar methods have been employed to study other modifications to \ac{gr} \cite{LIGOScientific:2019fpa}. 
In principle, we could also study higher-order CPT-violating effects on \acp{gw} in the same way.
However, we expect significantly weaker constraints on them \cite{Wang:2017igw}, which are left to future works.  
Furthermore, we should note that a multi-band observation may improve our results significantly, since the CPT-violating effect is proportional to the square of \ac{gw} frequency.
In addition, we expect CPT invariance to be further tested in the near future \cite{Wang:2017igw}, since more and more \acp{bbh} will be detected by upcoming observing runs of \ac{ligo} and Virgo and other observatories under construction \cite{Aasi:2013wya,Cornish:2018dyw}.

\begin{acknowledgements}\noindent
	This project is supported by National Natural Science Foundation of China  (Grant No. 12005016, No. 11690023 and No. 12021003), and by a grant upon Grant No. Y954040101 from the Institute of High Energy Physics, Chinese Academy of Sciences.
	We would like to appreciate Prof. Zhoujian Cao, Mr. Zucheng Chen and Mr. Shi-Chao Wu for helpful discussions and suggestions. 
\end{acknowledgements}


\begin{thebibliography}{10}
	
	
	\bibitem{Peskin:1995ev}
	M.~E.~Peskin and D.~V.~Schroeder,
	{\em An Introduction to quantum field theory},
	Addison-Wesley, Reading, USA (1995).
	
	\bibitem{Kostelecky:2008ts}
	V.~A.~Kostelecky and N.~Russell,
	``Data Tables for Lorentz and CPT Violation,''
	Rev.\ Mod.\ Phys.\  {\bf 83} (2011) 11,
	doi:10.1103/RevModPhys.83.11
	[arXiv:0801.0287 [hep-ph]].
	
	\bibitem{Will:2014kxa}
	C.~M.~Will,
	``The Confrontation between General Relativity and Experiment,''
	Living Rev.\ Rel.\  {\bf 17} (2014) 4,
	doi:10.12942/lrr-2014-4
	[arXiv:1403.7377 [gr-qc]].
	
	
	
	\bibitem{Toma:2012xa}
	K.~Toma {\it et al.},
	``Strict Limit on CPT Violation from Polarization of Gamma-Ray Burst,''
	Phys.\ Rev.\ Lett.\  {\bf 109} (2012) 241104,
	doi:10.1103/PhysRevLett.109.241104
	[arXiv:1208.5288 [astro-ph.HE]].
	
	\bibitem{Myers:2003fd}
	R.~C.~Myers and M.~Pospelov,
	``Ultraviolet modifications of dispersion relations in effective field theory,''
	Phys.\ Rev.\ Lett.\  {\bf 90} (2003) 211601,
	doi:10.1103/PhysRevLett.90.211601
	[arXiv:hep-ph/0301124].
	
	\bibitem{Amelino-Camelia:2016ohi}
	G.~Amelino-Camelia, G.~D'Amico, G.~Rosati and N.~Loret,
	``In-vacuo-dispersion features for GRB neutrinos and photons,''
	Nat.\ Astron.\  {\bf 1} (2017) 0139,
	doi:10.1038/s41550-017-0139
	[arXiv:1612.02765 [astro-ph.HE]].
	
	\bibitem{Jacob:2006gn}
	U.~Jacob and T.~Piran,
	``Neutrinos from gamma-ray bursts as a tool to explore quantum-gravity-induced Lorentz violation,''
	Nature Phys.\  {\bf 3} (2007) 87,
	doi:10.1038/nphys506
	[arXiv:hep-ph/0607145].
	
	\bibitem{Kostelecky:1988zi}
	V.~A.~Kostelecky and S.~Samuel,
	``Spontaneous Breaking of Lorentz Symmetry in String Theory,''
	Phys.\ Rev.\ D {\bf 39} (1989) 683,
	doi:10.1103/PhysRevD.39.683
	
	\bibitem{AmelinoCamelia:1997gz}
	G.~Amelino-Camelia, J.~R.~Ellis, N.~E.~Mavromatos, D.~V.~Nanopoulos and S.~Sarkar,
	``Tests of quantum gravity from observations of gamma-ray bursts,''
	Nature {\bf 393} (1998) 763,
	doi:10.1038/31647
	[arXiv:astro-ph/9712103].
	
	\bibitem{AmelinoCamelia:2008qg}
	G.~Amelino-Camelia,
	``Quantum-Spacetime Phenomenology,''
	Living Rev.\ Rel.\  {\bf 16} (2013) 5,
	doi:10.12942/lrr-2013-5
	[arXiv:0806.0339 [gr-qc]].
	
	\bibitem{Mielczarek:2017cdp}
	J.~Mielczarek and T.~Trześniewski,
	``Towards the map of quantum gravity,''
	Gen.\ Rel.\ Grav.\  {\bf 50} (2018) no.6,  68,
	doi:10.1007/s10714-018-2391-3
	[arXiv:1708.07445 [hep-th]].
	
	\bibitem{Abbott:2016blz}
	B.~P.~Abbott {\it et al.} [LIGO Scientific and Virgo Collaborations],
	``Observation of Gravitational Waves from a Binary Black Hole Merger,''
	Phys.\ Rev.\ Lett.\  {\bf 116} (2016) no.6,  061102,
	doi:10.1103/PhysRevLett.116.061102
	[arXiv:1602.03837 [gr-qc]].
	
	\bibitem{Isi:2019asy}
	M.~Isi, K.~Chatziioannou and W.~M.~Farr,
	``Hierarchical test of general relativity with gravitational waves,''
	Phys.\ Rev.\ Lett.\  {\bf 123} (2019) no.12,  121101,
	doi:10.1103/PhysRevLett.123.121101
	[arXiv:1904.08011 [gr-qc]].
	
	\bibitem{LIGOScientific:2019fpa}
	B.~P.~Abbott {\it et al.} [LIGO Scientific and Virgo Collaborations],
	``Tests of General Relativity with the Binary Black Hole Signals from the LIGO-Virgo Catalog GWTC-1,''
	Phys.\ Rev.\ D {\bf 100} (2019) no.10,  104036,
	doi:10.1103/PhysRevD.100.104036
	[arXiv:1903.04467 [gr-qc]].
	
	\bibitem{Isi:2019aib}
	M.~Isi, M.~Giesler, W.~M.~Farr, M.~A.~Scheel and S.~A.~Teukolsky,
	``Testing the no-hair theorem with GW150914,''
	Phys.\ Rev.\ Lett.\  {\bf 123} (2019) no.11,  111102,
	doi:10.1103/PhysRevLett.123.111102
	[arXiv:1905.00869 [gr-qc]].
	
	\bibitem{Yunes:2016jcc}
	N.~Yunes, K.~Yagi and F.~Pretorius,
	``Theoretical Physics Implications of the Binary Black-Hole Mergers GW150914 and GW151226,''
	Phys.\ Rev.\ D {\bf 94} (2016) no.8,  084002,
	doi:10.1103/PhysRevD.94.084002
	[arXiv:1603.08955 [gr-qc]].
	
	\bibitem{Addazi:2018uhd}
	A.~Addazi, A.~Marciano and N.~Yunes,
	``Can we probe Planckian corrections at the horizon scale with gravitational waves?,''
	Phys.\ Rev.\ Lett.\  {\bf 122} (2019) no.8,  081301,
	doi:10.1103/PhysRevLett.122.081301
	[arXiv:1810.10417 [gr-qc]].
	
	\bibitem{Wu:2016igi}
	X.~F.~Wu, H.~Gao, J.~J.~Wei, P.~Mészáros, B.~Zhang, Z.~G.~Dai, S.~N.~Zhang and Z.~H.~Zhu,
	``Testing Einstein’s weak equivalence principle with gravitational waves,''
	Phys.\ Rev.\ D {\bf 94} (2016) 024061,
	doi:10.1103/PhysRevD.94.024061
	[arXiv:1602.01566 [astro-ph.HE]].
	
	
	\bibitem{Cardoso:2017cqb}
	V.~Cardoso and P.~Pani,
	``Tests for the existence of black holes through gravitational wave echoes,''
	Nat.\ Astron.\  {\bf 1} (2017) no.9,  586,
	doi:10.1038/s41550-017-0225-y
	[arXiv:1709.01525 [gr-qc]].
	
	\bibitem{Ezquiaga:2018btd}
	J.~M.~Ezquiaga and M.~Zumalacárregui,
	``Dark Energy in light of Multi-Messenger Gravitational-Wave astronomy,''
	Front.\ Astron.\ Space Sci.\  {\bf 5} (2018) 44,
	doi:10.3389/fspas.2018.00044
	[arXiv:1807.09241 [astro-ph.CO]].
	
	\bibitem{Wang:2016ana}
	S.~Wang, Y.~F.~Wang, Q.~G.~Huang and T.~G.~F.~Li,
	``Constraints on the Primordial Black Hole Abundance from the First Advanced LIGO Observation Run Using the Stochastic Gravitational-Wave Background,''
	Phys. Rev. Lett. \textbf{120}, no.19, 191102 (2018),
	doi:10.1103/PhysRevLett.120.191102
	[arXiv:1610.08725 [astro-ph.CO]].
	
	\bibitem{Chang:2019xcb}
	Z.~Chang, Q.~G.~Huang, S.~Wang and Z.~C.~Zhao,
	``Low-redshift constraints on the Hubble constant from the baryon acoustic oscillation “standard rulers” and the gravitational wave “standard sirens”,''
	Eur. Phys. J. C \textbf{79}, no.2, 177 (2019),
	doi:10.1140/epjc/s10052-019-6664-0
	
	\bibitem{Zhao:2019xmm}
	W.~Zhao, T.~Zhu, J.~Qiao and A.~Wang,
	``Waveform of gravitational waves in the general parity-violating gravities,''
	Phys.\ Rev.\ D {\bf 101} (2020) no.2,  024002,
	doi:10.1103/PhysRevD.101.024002
	[arXiv:1909.10887 [gr-qc]].
	
	\bibitem{Kostelecky:2016kfm}
	V.~A.~Kostelecký and M.~Mewes,
	``Testing local Lorentz invariance with gravitational waves,''
	Phys.\ Lett.\ B {\bf 757} (2016) 510,
	doi:10.1016/j.physletb.2016.04.040
	[arXiv:1602.04782 [gr-qc]].
	
	
	
	
	\bibitem{Wang:2017igw}
	S.~Wang,
	``Exploring the CPT violation and birefringence of gravitational waves with ground- and space-based gravitational-wave interferometers,''
	Eur. Phys. J. C \textbf{80}, no.4, 342 (2020),
	doi:10.1140/epjc/s10052-020-7812-2
	[arXiv:1712.06072 [gr-qc]].
	
	\bibitem{Mewes:2019dhj}
	M.~Mewes,
	``Signals for Lorentz violation in gravitational waves,''
	Phys.\ Rev.\ D {\bf 99} (2019) no.10,  104062,
	doi:10.1103/PhysRevD.99.104062
	[arXiv:1905.00409 [gr-qc]].
	
	\bibitem{Kostelecky:2003fs}
	V.~A.~Kostelecky,
	``Gravity, Lorentz violation, and the standard model,''
	Phys.\ Rev.\ D {\bf 69} (2004) 105009,
	doi:10.1103/PhysRevD.69.105009
	[hep-th/0312310].
	
	\bibitem{Bailey:2014bta}
	Q.~G.~Bailey, A.~Kostelecký and R.~Xu,
	``Short-range gravity and Lorentz violation,''
	Phys.\ Rev.\ D {\bf 91} (2015) no.2,  022006,
	doi:10.1103/PhysRevD.91.022006
	[arXiv:1410.6162 [gr-qc]].
	
	
	\bibitem{Thrane:2018qnx}
	E.~Thrane and C.~Talbot,
	``An introduction to Bayesian inference in gravitational-wave astronomy: parameter estimation, model selection, and hierarchical models,''
	Publ.\ Astron.\ Soc.\ Austral.\  {\bf 36} (2019) e010,
	doi:10.1017/pasa.2019.2
	[arXiv:1809.02293 [astro-ph.IM]].
	
	\bibitem{Maggiore:1900zz}
	M.~Maggiore, {\em {Gravitational Waves. Vol. 1: Theory and Experiments}}, Oxford Master Series in Physics, Oxford University Press (2007).
	
	\bibitem{Lehner:2001wq}
	L.~Lehner,
	``Numerical relativity: A Review,''
	Class.\ Quant.\ Grav.\  {\bf 18} (2001) R25,
	doi:10.1088/0264-9381/18/17/202
	[gr-qc/0106072].
	
	
	
	
	\bibitem{Hannam:2013oca}
	M.~Hannam, P.~Schmidt, A.~Bohé, L.~Haegel, S.~Husa, F.~Ohme, G.~Pratten and M.~Pürrer,
	``Simple Model of Complete Precessing Black-Hole-Binary Gravitational Waveforms,''
	Phys.\ Rev.\ Lett.\  {\bf 113} (2014) no.15,  151101,
	doi:10.1103/PhysRevLett.113.151101
	[arXiv:1308.3271 [gr-qc]].
	
	\bibitem{Schmidt:2014iyl}
	P.~Schmidt, F.~Ohme and M.~Hannam,
	``Towards models of gravitational waveforms from generic binaries II: Modelling precession effects with a single effective precession parameter,''
	Phys.\ Rev.\ D {\bf 91} (2015) no.2,  024043,
	doi:10.1103/PhysRevD.91.024043
	[arXiv:1408.1810 [gr-qc]].
	
	\bibitem{LSC-Virgo-2019}
	LIGO Scientific and Virgo Collaborations, ``The LSC-Virgo White Paper on Gravitational Wave Data Analysis and Astrophysics'', LIGO Document LIGO-T1900541-v2 (2019). https://dcc.ligo.org/LIGO-T1900541/public
	
	
\bibitem{Smith:2019ucc}
R.~Smith, G.~Ashton, A.~Vajpeyi and C.~Talbot,
``Massively parallel Bayesian inference for transient gravitational-wave astronomy,''
[arXiv:1909.11873 [gr-qc]].
	
	
	
	
	\bibitem{LIGOScientific:2018mvr}
	B.~P.~Abbott {\it et al.} [LIGO Scientific and Virgo Collaborations],
	``GWTC-1: A Gravitational-Wave Transient Catalog of Compact Binary Mergers Observed by LIGO and Virgo during the First and Second Observing Runs,''
	Phys.\ Rev.\ X {\bf 9} (2019) no.3,  031040,
	doi:10.1103/PhysRevX.9.031040
	[arXiv:1811.12907 [astro-ph.HE]].
	
	\bibitem{Abbott:2016nmj}
	B.~P.~Abbott {\it et al.} [LIGO Scientific and Virgo Collaborations],
	``GW151226: Observation of Gravitational Waves from a 22-Solar-Mass Binary Black Hole Coalescence,''
	Phys.\ Rev.\ Lett.\  {\bf 116} (2016) no.24,  241103,
	doi:10.1103/PhysRevLett.116.241103
	[arXiv:1606.04855 [gr-qc]].
	
	\bibitem{Audren:2012wb}
	B.~Audren, J.~Lesgourgues, K.~ Benabed, and S.~Prunet,
	``Conservative Constraints on Early Cosmology: an illustration of the Monte Python cosmological parameter inference code,''
	JCAP 1302  (2013) 001,
	doi:10.1088/1475-7516/2013/02/001
	[arXiv:1210.7183[astro-ph.CO]].
	
\bibitem{Shao:2020shv} 
  L.~Shao,
  ``Combined search for anisotropic birefringence in the gravitational-wave transient catalog GWTC-1,''
  Phys.\ Rev.\ D {\bf 101}, no. 10, 104019 (2020)
  doi:10.1103/PhysRevD.101.104019
  [arXiv:2002.01185 [hep-ph]].
	
	\bibitem{Aasi:2013wya}
	B.~P.~Abbott {\it et al.} [KAGRA and LIGO Scientific and VIRGO Collaborations],
	``Prospects for Observing and Localizing Gravitational-Wave Transients with Advanced LIGO, Advanced Virgo and KAGRA,''
	Living Rev.\ Rel.\  {\bf 21} (2018) no.1,  3,
	doi:10.1007/s41114-018-0012-9, 10.1007/lrr-2016-1
	[arXiv:1304.0670 [gr-qc]].
	
	\bibitem{Cornish:2018dyw}
	T.~Robson, N.~J.~Cornish and C.~Liu,
	``The construction and use of LISA sensitivity curves,''
	Class.\ Quant.\ Grav.\  {\bf 36} (2019) no.10,  105011,
	doi:10.1088/1361-6382/ab1101
	[arXiv:1803.01944 [astro-ph.HE]].
	
	
	
\end{thebibliography}

\end{document}